\documentclass[11pt]{article}
\usepackage{amssymb}
\usepackage{amsmath}
\usepackage{amstext}
\usepackage{graphicx,epsfig}
\usepackage{epsfig}
\usepackage{verbatim} 
\usepackage{fancybox}
\usepackage{color}
\usepackage{ulem}
\usepackage{enumitem}
\usepackage{subfigure}
\usepackage{bbm}
\usepackage{parskip}
\usepackage{dsfont}
\usepackage[numbers,sort&compress]{natbib}

\newcommand{\Comment}[1]{{}}
\definecolor{darkblue}{rgb}{0.15,0.35,0.55}
\definecolor{reddish}{rgb}{0.65, 0.2, 0.2}
\usepackage[linktocpage=true]{hyperref}
\hypersetup{
colorlinks=true,
citecolor=darkblue,
linkcolor=reddish,
urlcolor=darkblue,
pdfauthor={},
pdftitle={},
pdfsubject={}
}

\newcommand{\be}{\begin{equation}}
\newcommand{\ee}{\end{equation}}
\newcommand{\bea}{\begin{eqnarray}}
\newcommand{\eea}{\end{eqnarray}}
\newcommand{\beas}{\begin{eqnarray*}}
\newcommand{\eeas}{\end{eqnarray*}}
\newcommand{\nn}{\nonumber \\ }

\def\({\left(}
\def\){\right)}

\newcommand{\rd}{{\rm d}}

\newcommand{\half}{\frac{1}{2}}

\def\gsim{ \lower .75ex \hbox{$\sim$} \llap{\raise .27ex \hbox{$>$}} }
\def\lsim{ \lower .75ex \hbox{$\sim$} \llap{\raise .27ex \hbox{$<$}} }

\title{}
\author{}

\numberwithin{equation}{section}

\begin{document}
~
\vspace{1truecm}
\renewcommand{\thefootnote}{\fnsymbol{footnote}}
\begin{center}
{\huge \bf{Constructing Galileons}}
\end{center} 

\vspace{1truecm}
\thispagestyle{empty}
\centerline{\Large Mark Trodden\footnote{\tt trodden@physics.upenn.edu}}
\vspace{.7cm}

\centerline{\it Center for Particle Cosmology, Department of Physics and Astronomy,}
\centerline{\it University of Pennsylvania, Philadelphia, PA 19104, USA}

\vspace{.5cm}
\begin{abstract}
\vspace{.03cm}
\noindent
In this plenary talk delivered at the DISCRETE 2014 conference in London, I briefly summarize the ideas behind and attractive properties of the Galileon field theories, and describe a broad class of scalar field theories that share these properties. After describing how Galileons arise, and commenting on their fascinating properties, in the latter half of the talk I focus on novel ways of constructing Galileon-like theories, using both the probe brane construction, and the coset construction.
\end{abstract}

\newpage

\tableofcontents

\newpage
\renewcommand*{\thefootnote}{\arabic{footnote}}
\setcounter{footnote}{0}

\section{Introduction} 
The problem of cosmic acceleration has prompted the search for both new and well-behaved forms of dark energy and ghost free infrared modifications of General Relativity (GR). At the level of effective field theory (EFT), around a given background, either approach boils down to a field theory of interacting fields, about which we may ask a variety of questions of theoretical consistency (and, of course, observational viability). These include requirements that the theory be ghost-free, that it remain in the weakly-coupled, calculable regime in the environments we are interested in, and that the theory be technically natural. There are also more technical questions that the theory may face, such as whether there are any obstacles to a Lorentz-invariant ultraviolet (UV) completion, arising, for example, from certain analyticity properties of the S-matrix.

These requirements, coupled with those from demanding that any new long-range force of gravitational strength not ruin precision tests of GR within the solar system and local universe, have proven extremely restrictive, with very few new ideas remaining viable. On the other hand, this search has yielded a variety of fascinating new ideas that remain of interest for cosmic acceleration, but also present intriguing new theoretical challenges all on their own. The more intriguing of these ideas have arisen from studies of extra-dimensional brane-localized gravity theories~\cite{Dvali:2000hr}, and ghost free massive gravity~\cite{deRham:2010ik,deRham:2010kj}. In both cases, there exists a limit in which a fascinating scalar field theory with higher-derivative interaction results - the {\it Galileons}.

In this summary of a plenary lecture, delivered at the DISCRETE 2014 conference, I briefly introduce the idea of Galileons, their attractive properties and their interesting features. I then describe how one may construct theories such as these from two different viewpoints. The first is the probe brane construction, which I show can be generalized to yield new theories, and the second is the coset construction, which can she new light on the topological properties of such theories. As is usual in a proceedings, I will have space only to reference those papers of most direct relevance to my own thinking on the subject, and will not attempt to be comprehensive.

\section{Galileons, the DGP Model and Massive Gravity}
An easy starting point from which to motivate Galileons is the Dvali-Gabadadze-Porrati (DGP) model~\cite{Dvali:2000hr}, in which a $3+1$-dimensional brane is embedded in a $5$d bulk, with action
\be
S=\frac{M_5^3}{2}\int d^5X\, \sqrt{-G}\,\  R[G] + \frac{M_4^2}{2}\int d^4x\, \sqrt{-g}\, \ R[g] \ .
\label{DGPaction}
\ee
In this model gravity is modified on large distances, and there exists a branch of cosmological solutions that exhibit self-acceleration at late times, although this branch contains a ghost. There also exists a ghost-free ``normal" branch, for which a 4d effective action can be derived which, in a particular ``decoupling" limit, reduces to a theory of a single scalar $\pi$, with a cubic self-interaction term $\sim (\partial\pi)^2\Box \pi$.  This term yields second order field equations and is invariant (up to a total derivative) under the {\it Galilean} transformations
\be 
\label{Galileoninvarianceold}
\pi (x) \rightarrow \pi(x) + c + b_\mu x^\mu \ ,
\ee
with $c$ and $b_{\mu}$ constants.  

It is natural to ask what other terms are allowed in a four dimensional field theory with this same symmetry~\cite{Nicolis:2008in}. Referring to the field of such a theory as the {\it Galileon}, it turns out that, in addition to an infinite number of terms in which two derivatives act on every field (trivially invariant under the galilean symmetry) there are a finite number of terms that have fewer numbers of derivatives per field. These possess second order equations of motion, and there can exist regimes in which these special terms are important, and the infinity of other possible terms within the effective field theory are not. Along with a non-renormalization theorem for Galileons~\cite{Luty:2003vm,Hinterbichler:2010xn,Burrage:2010cu}, this means that we are able, in some circumstances, to understand non-linear effects that are quantum mechanically exact.

For $n\geq 1$, the $(n+1)$-th order Galileon Lagrangian (where ``order" refers to the number of copies of the scalar field $\pi$ that appear in the term) is
\be
\label{Galileon2} 
{\cal L}_{n+1}=n\eta^{\mu_1\nu_1\mu_2\nu_2\cdots\mu_n\nu_n}\left( \partial_{\mu_1}\pi\partial_{\nu_1}\pi\partial_{\mu_2}\partial_{\nu_2}\pi\cdots\partial_{\mu_n}\partial_{\nu_n}\pi\right) \ ,
\ee 
where 
\be
\label{tensor} 
\eta^{\mu_1\nu_1\mu_2\nu_2\cdots\mu_n\nu_n}\equiv{1\over n!}\sum_p\left(-1\right)^{p}\eta^{\mu_1p(\nu_1)}\eta^{\mu_2p(\nu_2)}\cdots\eta^{\mu_np(\nu_n)} 
\ee 
and the sum is over all permutations of the $\nu$ indices, with $(-1)^p$ the sign of the permutation.  The tensor~(\ref{tensor}) is antisymmetric in the $\mu$ indices, antisymmetric in the $\nu$ indices, and symmetric under interchange of any $\mu,\nu$ pair with any other.  These Lagrangians are unique up to total derivatives and overall constants.   Because of the antisymmetry requirement on $\eta$, only the first $n$ of these Galileons are non-trivial in $n$-dimensions.  In addition, the tadpole term, $\pi$, is Galilean invariant, and we therefore include it as the first-order Galileon.

These Galileon actions can be generalized to the multi-field case, where there is a multiplet $\pi^I$ of fields~\cite{Deffayet:2010zh,Padilla:2010de,Padilla:2010ir,Hinterbichler:2010xn,Zhou:2010di,Padilla:2010tj}. It should be remembered that all these theories are not renormalizable, and should be considered as effective field theories with some cutoff $\Lambda$, above which some UV completion is required.   However, as I have mentioned, there can exist regimes in which the above terms dominate, since the symmetries forbid any renormalizable terms, and other terms will have more derivatives.   One important consequence of these properties is that galileon theories exhibit a particular type of screening mechanism - the Vainshtein mechanism~\cite{Vainshtein:1972sx} - that allows them to mediate long range forces, competing with gravity at cosmic scales, while having negligible physical effects at the scales of local tests of gravity. This is the key to their use in modified gravity applications (see~\cite{Joyce:2014kja} for a recent review of these mechanisms and their effects and tests).

\section{The Probe Brane Construction and its Generalization}
One way to construct the Galileons is to generalize the DGP model to a general action for a probe brane floating in a higher dimensional flat bulk, and then to take a limit in which only a scalar field degree of freedom remains~\cite{deRham:2010eu}.  This {\it probe brane} construction can then be generalized in a variety of ways~\cite{Goon:2011qf,Goon:2011uw,Goon:2011xf,Goon:2012mu,Burrage:2011bt} to generate new galileon-like theories possessing many of the attractive features of the original model.\footnote{See~\cite{Trodden:2011xh} for a more comprehensive review of this.}

One generalization is to introduce additional co-dimensions into the construction. One can then show that the resulting theory is a $4$-d field theory of multiple scalar fields with a single possible interaction term, and thus a single free coupling constant. As for the original galileons,  this term is not renormalized to any order in perturbation theory~\cite{Hinterbichler:2010xn}.

Another generalization of the probe brane construction can be obtained~\cite{Goon:2011qf,Goon:2011uw,Goon:2011xf} by relaxing the assumption of flat geometry in the original method. To see this, consider a dynamical 3-brane in a fixed (4+1)-dimensional background, but now with a general bulk metric.  The dynamical variables are the brane embedding $X^A(x)$, which are five functions of the world-volume coordinates $x^\mu$.  The bulk metric $G_{AB}(X)$ allows us to construct the induced metric $\bar g_{\mu\nu}(x)$ and the extrinsic curvature $K_{\mu\nu}(x)$ via
\bea 
\bar g_{\mu\nu}&=&e^A_{\ \mu}e^B_{\ \nu} G_{AB}(X) \ , \\ 
K_{\mu\nu}&=&e^A_{\ \mu}e^B_{\ \nu}\nabla_A n_B \ ,
\eea
where $n^A$ is the unit normal vector, and $e^A_{\ \mu}= {\partial X^A\over\partial x^\mu}$ are the tangent vectors to the brane.

To ensure that the world-volume action is gauge invariant under brane reparametrizations, we write the action as a diffeomorphism scalar, $F$, of $\bar g_{\mu\nu}$, $K_{\mu\nu}$, the covariant derivative $\bar\nabla_\mu$ and the curvature $\bar R^\alpha_{\ \beta\mu\nu}$ constructed from $\bar g_{\mu\nu}$.
\be
\label{generalaction} 
S= \int d^4x\ \sqrt{-\bar g}F\left(\bar g_{\mu\nu},\bar\nabla_\mu,\bar R^{\alpha}_{\ \beta\mu\nu},K_{\mu\nu}\right) \ .
\ee
Requiring that the resulting equations of motion are second order, and hence avoid obvious ghosts, the allowed actions turn out to be the Lovelock~\cite{Lovelock:1971yv} and boundary~\cite{Gibbons:1976ue,York:1972sj,Myers:1987yn} terms.

Now, global symmetries of this action exist only if the bulk metric has Killing symmetries. To exploit this, we choose the gauge
\be
\label{physgauge} 
X^\mu(x)=x^\mu, \ \ \ X^5(x)\equiv \pi(x) \ ,
\ee
such that the bulk is foliated by time-like slices given by the surfaces $X^5(x)= {\rm constant}$. This completely fixes the gauge symmetry, and the leaves of the foliation are then parametrized by the arbitrary choices of the remaining coordinates $X^\mu$. The coordinate $\pi(x)$ measures the transverse position of the brane relative to the foliation in this gauge, and the resulting action for $\pi$ is
\be
\label{gaugefixedaction} 
S= \int d^4x\ \left. \sqrt{-\bar g}F\left(\bar g_{\mu\nu},\bar\nabla_\mu,\bar R^{\alpha}_{\ \beta\mu\nu},K_{\mu\nu}\right)\right|_{X^\mu=x^\mu,\ X^5=\pi} \ .
\ee

Any global symmetries of~(\ref{generalaction}) are global symmetries of~(\ref{gaugefixedaction}), but their form depends on the gauge. If a Killing vector $K^A$ generates a transformation, then we may restore the gauge~(\ref{physgauge}) by making a compensating gauge transformation $\delta_{g,{\rm comp}}x^\mu=-K^\mu$.  These two symmetries then combine to yield the following transformation of $\pi$
\be
\label{gaugefixsym} 
(\delta_K+\delta_{g,{\rm comp}})\pi=-K^\mu(x,\pi)\partial_\mu\pi+K^5(x,\pi) \ ,
\ee
which is a symmetry of the gauge fixed action~(\ref{gaugefixedaction}).

Simple, yet illuminating examples can be obtained by making a few assumptions. We assume our foliation is Gaussian normal with respect to $G_{AB}$, and we further demand that the extrinsic curvature on each slice be proportional to the induced metric. The metric then takes the form
\be 
\label{metricform} 
G_{AB}dX^AdX^B=d\rho^2+f(\rho)^2g_{\mu\nu}(x)dx^\mu dx^\nu \ ,
\ee
where $X^5=\rho$ denotes the transverse coordinate, and $g_{\mu\nu}(x)$ is an arbitrary brane metric.  This special case includes all examples in which a maximally symmetric ambient space is foliated by maximally symmetric slices.  Killing symmetries which preserve the foliation will be linearly realized, whereas those that do not are realized nonlinearly.  Thus, the algebra of all Killing vectors is spontaneously broken to the subalgebra of Killing vectors preserving the foliation.

In the gauge (\ref{physgauge}), the induced metric is $\bar g_{\mu\nu}=f(\pi)^2g_{\mu\nu}+\nabla_\mu\pi\nabla_\nu\pi$.
Defining the quantity $\gamma=1/ \sqrt{1+{1\over f^2}(\nabla\pi)^2}$,
the extrinsic curvature is then
\be 
K_{\mu\nu}=\gamma\left(-\nabla_\mu\nabla_\nu\pi+f f'g_{\mu\nu}+2{f'\over f}\nabla_\mu\pi\nabla_\nu\pi\right) \ .
\ee

On the 4-dimensional brane, as described, we can add four Lovelock and boundary terms, plus a tadpole term,
\bea   {\cal L}_1&=&\sqrt{-g}\int^\pi d\pi' f(\pi')^4,\nn\\
{\cal L}_2&=&- \sqrt{-\bar g} \ ,\nn\\
{\cal L}_3&=& \sqrt{-\bar g}K \ ,\nn\\
{\cal L}_4&=& -\sqrt{-\bar g}\bar R \ ,\nn\\
{\cal L}_5&=&{3\over 2}\sqrt{-\bar g} {\cal K}_{\rm GB} \ ,
\label{ghostfreegenterms} \eea
where the explicit form of the Gauss-Bonnet boundary term is
${\cal K}_{\rm GB}=-{1\over3}K^3+K_{\mu\nu}^2K-{2\over 3}K_{\mu\nu}^3-2\left(\bar R_{\mu\nu}-\half \bar R \bar g_{\mu\nu}\right)K^{\mu\nu}$. 

$\mathcal L_1$ is the zero derivative tadpole term, which is the proper volume between any $\rho=$ constant surface and the brane position, $\pi(x)$ \cite{Goon:2011qf}.  While different in origin from the other terms, it too has the symmetry~(\ref{gaugefixsym}).  Each of these terms may appear in a general Lagrangian with an arbitrary coefficient. 

Evaluating these expressions for the metric~(\ref{metricform}) involves a lengthy calculation which ultimately yields
\bea   
{\cal L}_1&=&\sqrt{-g}\int^\pi d\pi' f(\pi')^4,\nn\\
{\cal L}_2&=&-\sqrt{-g}f^4\sqrt{1+{1\over f^2}(\partial\pi)^2},\nn\\
{\cal L}_3&=&\sqrt{-g}\left[f^3f'(5-\gamma^2)-f^2[\Pi]+\gamma^2[\pi^3]\right],\nn \\
{\cal L}_4&=& -\sqrt{-g}\Big\{{1\over\gamma}f^2R-2{\gamma}R_{\mu\nu}\nabla^\mu\pi\nabla^\nu\pi \nn\\
&&\ \ +\gamma\left[[\Pi]^2-[\Pi^2]+2{\gamma^2\over f^2}\left(-[\Pi][\pi^3]+[\pi^4]\right)\right]\nn\\
&&\ \ \ \!+6{f^3f''\over \gamma}\left(-1+\gamma^2\right) \!	 \nn \\
&&\ \ \ \ +2\gamma ff'\left[-4[\Pi]+{\gamma^2\over f^2}\left(f^2[\Pi]+4[\pi^3]\right)\right]\nn\\
&&\ \ \ \ \ -6{f^2f'^2\over \gamma}\left(1-2\gamma^2+\gamma^4\right) \Big\} \ ,
\eea
where the expression for ${\cal L}_5$ is too long to write here, but its full form is given in~\cite{Goon:2011qf}. In these expressions, all curvatures and covariant derivatives are those of the background metric $g_{\mu\nu}$, and $\Pi$ denotes the matrix of covariant second dervatives $\Pi_{\mu\nu}\equiv\nabla_{\mu}\nabla_\nu\pi$. We use the notation $[\Pi^n]\equiv Tr(\Pi^n)$, e.g. $[\Pi]=\square\pi$, $[\Pi^2]=\nabla_\mu\nabla_\nu\pi\nabla^\mu\nabla^\nu\pi$, and $[\pi^n]\equiv \nabla\pi\cdot\Pi^{n-2}\cdot\nabla\pi$, i.e. $[\pi^2]=\nabla_\mu\pi\nabla^\mu\pi$, $[\pi^3]=\nabla_\mu\pi\nabla^\mu\nabla^\nu\pi\nabla_\nu\pi$. The equations of motion contain no more than two derivatives on each field, ensuring that no extra degrees of freedom propagate around any background.

There are interesting cases in which the 5d background metric has 15 global symmetries, the maximal number -- the bulk can be either $5$d anti-de Sitter space $AdS_5$ with isometry algebra $so(4,2)$, 5d de-Sitter space $dS_5$ with isometry algebra $so(5,1)$, or flat 5d Minkowski space $M_5$ with isometry algebra the five dimensional Poincar{\'e} algebra $p(4,1)$.  In addition, there are cases in which the brane metric $g_{\mu\nu}$, and hence the extrinsic curvature, are maximally symmetric, so that the unbroken subalgebra has the maximal number of generators, 10.  This means that the leaves of the foliation are either $4$d anti-de Sitter space $AdS_4$ with isometry algebra $so(3,2)$, 4d de-Sitter space $dS_4$ with isometry algebra $so(4,1)$, or flat 4d Minkowski space $M_4$ with isometry algebra $p(3,1)$.  In fact, there are only 6 such possible foliations of $5$d maximally symmetric spaces by $4$d maximally symmetric time-like slices, such that the metric takes the form~(\ref{metricform}).  Flat $M_5$ can be foliated by flat $M_4$ slices or by $dS_4$ slices; $dS_5$ can be foliated by flat $M_4$ slices, $dS_4$ slices, or $AdS_4$ slices; and $AdS_5$ can only be foliated by $AdS_4$ slices.  Each of these 6 foliations, through the construction leading to~(\ref{gaugefixedaction}), will generate a class of theories living on an $AdS_4$, $M_4$ or $dS_4$ background and having 15 global symmetries broken to the 10 isometries of the brane, the same numbers as those of the original galileon.  

To see analogues of the galileons emerge, we expand the Lagrangians in powers of $\lambda$ around some constant background, $\pi\rightarrow\pi_0+\lambda\pi$.  There exist linear combinations of the Lagrangians, $\bar{\mathcal L}_n=c_1\mathcal L_1+\ldots +c_n\mathcal L_n$, for which all terms $\mathcal{O}\left (\lambda^{n-1}\right )$ or lower are total derivatives.  Carrying this out for the remaining four maximally symmetric cases in which the 4d background is curved, new classes of theories are produced, and canonically normalizing via $\hat\pi={1\over L^2}\pi$, where $L$ is the $dS_4$ or $AdS_4$ radius, the Lagrangians become
\begin{eqnarray} 
\hat{\cal L}_1&=&\sqrt{-g}\hat\pi \ , \nn\\
\hat{\cal L}_2&=&-\half\sqrt{-g} \left((\partial\hat\pi)^2-{R\over 3}\hat \pi^2\right) \ ,\nn \\
\hat{\cal L}_3&=& \sqrt{-g}\left(-{1\over 2}(\partial\hat\pi)^2[\hat\Pi]-{R\over 4} (\partial\hat\pi)^2\hat\pi+{R^2\over 36}\hat\pi^3\right) \ ,\nn\\
\hat{\cal L}_4&=&\sqrt{-g}\Big[-\half(\partial\hat\pi)^2\Big([\hat\Pi]^2-[\hat\Pi^2]+{R\over 24}(\partial\hat\pi)^2\nn\\
&&\ \ \ \ \ \ \ \ \  \ \ \ +{R\over 2}\hat\pi[\hat\Pi]+{R^2\over 8}\hat\pi^2\Big)+{R^3\over 288}\hat\pi^4\Big] \ ,
\label{singlesetGalileons} 
\eea 
with ${\cal L}_5$ again found in~\cite{Goon:2011qf}.

In the above $R=\pm{12\over L^2}$ is the Ricci curvature of the $dS_4$ or $AdS_4$ background. These Lagrangians describe Galileons that propagate on curved space yet retain the same number of symmetries as the full theory, whose form comes from expanding~(\ref{gaugefixsym}) in appropriate powers of $\lambda$.  For example, in the case of a $dS_4$ background in conformal inflationary coordinates $(u,y^i)$, the non-linear symmetries are
\be \label{dSGalileontrans}
\delta_{+}\hat\pi={1\over u}\left(u^2-y^2\right) , \ \ \
\delta_{-} \hat\pi=-{1\over u},\ \ \ 
\delta_{i} \hat\pi = {y_i\over u} \ .
\ee

One interesting feature of these models which is not true for the flat space Galileon theories is the existence of potentials with couplings determined by the symmetries~(\ref{gaugefixsym}).  Specifically, the scalar field acquires a mass of order the $dS_4$ or $AdS_4$ radius, with a value protected by the symmetries (\ref{dSGalileontrans}). This means that small masses can be made technically natural, which can be handy for cosmological applications.

The methods presented in this section can be used to create large numbers of galileon-like models, with potentially extremely complex symmetries, which would be hard to discover in other ways.

\section{The Coset Construction and Wess-Zumino Terms}
A different, and complementary, way to investigate galileon and galileon-like theories is to take advantage of the fact that the relevant symmetries are not realized linearly on the fields. For example, the basic galilean symmetry, $\pi \rightarrow \pi + c +b_{\mu}x^{\mu}$, is a combination of two transformations - a shift symmetry and a shift symmetry acting on the derivative, neither of which is linearly realized on the field $\pi$. One way to think about nonlinearly realized symmetries is in the language of spontaneous symmetry breaking. If we have a symmetry group, $G$, broken down to a subgroup $H$, then generically the broken phase linearly realizes the preserved  subgroup and nonlinearly realizes the elements $G/H$ that are no longer preserved. Given the breaking pattern, there exists an algorithmic method - the ``coset construction"~\cite{Coleman:1969sm,Callan:1969sn,volkov,xthschool,inversehiggs} - for constructing the most general Lagrangians linearly realizing $H$ and nonlinearly realizing the remaining symmetries. The low energy physics of a spontaneously broken system is described by the Goldstone modes that result from this construction. 

The coset construction can be used even when the dynamics leading to symmetry breaking are unknown or not understood, such as in the classic application to the physics of pions~\cite{Weinberg:1968de}. Furthermore, this construction can be applied to gauge theories by noticing that they nonlinearly realize the local versions of their symmetry groups, and this application can be extended to the physics of such theories in the Higgs phase~\cite{Goon:2014ika}.  
 
Consider a Lie group $G$, broken down to a subgroup $H$, and represent the generators of $H$ by $\{V_{I}\}$, and those of the remaining -- broken -- generators by $\{Z_{a}\}$. Assuming that the commutator of an element of $\{V_{I}\}$ with an element of $\{Z_{a}\}$ does not contain another generator from the $\{V_{I}\}$, then we may write a  canonical representative element of the coset $G/H$ as $g(\xi)\equiv \exp \left(\xi^{a}Z_{a}\right)$, where the Goldstone fields are represented by the $\xi^{a}$.  

Any element $g'\in G$ generates a transformation $g(\xi)\mapsto\tilde{g}(\xi,g')$, where we define $\tilde{g}(\xi,g')$ by
 \begin{align}
 g'g(\xi)=\tilde{g}(\xi,g')h(\xi,g') \ ,
 \label{MCtransformationcondition}
 \end{align}  
where $h(\xi,g)\in H$. In this manner, as required, the Goldstone fields linearly realize the symmetries of the preserved subgroup $H$ and nonlinearly realize the remaining broken symmetries. This can be seen by writing $\tilde{g}(\xi,g')\equiv \exp(\tilde\xi^{a}Z_{a} )$, and noting that the relation between $\tilde\xi^{a}$ and $\xi^{a}$ is linear if $g' \in H$, but  nonlinear otherwise.  
 
The first step in constructing actions is to build a Lie algebra-valued 1-form known as the Maurer--Cartan (MC) form $\Omega\equiv g^{-1}(\xi)\rd g(\xi)$.  Decomposing this into its parts along the broken and unbroken generators as
 \begin{align}
 \Omega&\equiv \Omega^{a}Z_{a}+\Omega^{I}V_{I}\equiv \Omega_{Z}+\Omega_{V}\ ,
 \end{align}
one sees that an arbitrary element $g'\in G$ induces a transformation under which the components of the MC form transform as 
\begin{align}
 g':\begin{cases} \Omega_{Z}&\longmapsto ~h(\xi,g')\Omega_{Z}h^{-1}(\xi,g'),\\
\Omega_{V}&\longmapsto ~ h(\xi,g')\left (\Omega_{V}+\rd\right )h^{-1}(\xi,g').\end{cases}\ \label{MCTransformation}
 \end{align}
These transformation properties under the action of the group $G$ are what make the MC form so useful. To build Lagrangians that are invariant under \textit{all} the symmetries of $G$ we combine factors of $\Omega_{Z}$ together and trace over group indices.  The operator that results is invariant under~\eqref{MCTransformation}, {\it i.e.}, it is $H$-invariant.  We then couple the Goldstone fields to other matter fields that transform in some representation of $H$ using the $\Omega_{V}$ components, which transform as a connection. 

At this point, we must deal with an issue that arises whenever spacetime symmetries are among those that are nonlinearly realized - the removal of fields via inverse Higgs (IH) constraints~\cite{Ivanov:1975zq}. Na\"ively, for a breaking pattern $G\to H$, we have $\dim(G/H)$ fields $\{\xi^{a}\}$ in the representative element $g$, which is the appropriate number of Goldstone modes for the case of internal symmetry breaking. However, this counting does not always work when spacetime symmetries are involved~\cite{Low:2001bw}.  In practice, if the commutator between a preserved translation generator, $P_{\mu}$, and a broken generator, $Z_{1}$, contains a second broken generator, $Z_{2}$, then we can set some part of the MC component along $Z_{2}$ to zero, and hence eliminate the Goldstone field corresponding to $Z_{1}$. These inverse Higgs constraints allow us to consistently reduce the number of fields while still realizing all the symmetries of $G$.

Another subtlety concerns how we deal with any preserved translation generators, $P_{\mu}$. When treating a case of spacetime symmetry breaking, we treat these generators on the same footing as the broken generators, since translations are nonlinearly realized on the spacetime coordinates. Taking this into account, the MC form is thus written as
\be
\Omega =\Omega^{a}Z_{a}+\Omega^{I}V_{I}+\Omega^{\mu}P_{\mu}~.
\ee
We may construct invariant actions by combining forms using the wedge product, and we may also form a covariant derivative from these objects. 
The components along $P_{\mu}$ define a vielbein, $e_{\nu}{}^{\mu}$, via $\Omega^{\mu}P_{\mu}\equiv \rd x^{\nu}e_{\nu}{}^{\mu}P_{\mu}$, which may be used to define an invariant measure, and define a covariant derivative of the Goldstone fields via $\Omega^{a}Z_{a}=\rd x^{\nu}e_{\nu}{}^{\mu}\mathcal{D}_{\mu}\xi^{a}Z_{a}$.  Terms in the invariant actions are then formed by contracting factors of covariant derivatives into $H$-invariant combinations, and finally integrating with the invariant measure.

So what happens when we try to use the coset construction for galileons in four dimensions? To start, recall that the galileons nonlinearly realize the following symmetries
\bea \delta_{C}\pi=1~,~~~~~~~~~~~~~~~~~~~~~~~~~\delta_{B^{\mu}}\pi=x^\mu \ ,
\eea
with non-trivial commutators
\bea
\left [P_{\mu },B_{\nu  }\right ] = \eta_{\mu \nu }C \ ,\ \ \ \ ~~~~~~~ \left[J_{\rho \sigma},B_{\nu }\right]=\eta_{\rho \nu }B_{\sigma }-\eta_{\sigma\nu }B_{ \rho} \ .
\eea
Along with Poincar\'e transformations, these comprise the galileon algebra $\mathfrak{Gal}(3+1,1)$, and so in 4$d$ galileons nonlinearly realize the breaking pattern
\be
\mathfrak{Gal}(3+1,1) \longrightarrow \mathfrak{iso}(3,1)~.
\ee
The coset is therefore parameterized by
\be
\tilde V = e^{x\cdot P}e^{\pi C+\xi\cdot B}~.
\ee
Because the linearly realized generators are the Lorentz transformations, the coset is
\be \label{cosetgal01}
{\rm Gal}(3+1,1)/{\rm SO}(3,1) \ .
\ee

The coefficients of the components of the MC form are
\be
\omega_P^\mu = \rd x^\mu~,~~~~~~~~~~~~
\omega_C = \rd\pi+\xi_\mu\rd x^\mu~,~~~~~~~~~~~~
\omega_B^\mu = \rd\xi^\mu~,~~~~~~~~~~~~
\omega_J^{\mu\nu} = 0~.~~~~~~~~~~~~
\label{singlefieldMCform}
\ee
Although there are two broken generators, $V_\mu$ and $C$, we only have a single Goldstone mode $\pi$ because of an inverse Higgs constraint -- the commutator $\left[P_\mu, B_\nu\right] = \eta_{\mu\nu} C$ implies that we can eliminate the $\xi_\mu$ field by setting $\omega_C = 0$, yielding
\be
\xi_\mu = -\partial_\mu\pi \ .
\ee
The components of the Maurer--Cartan form can then be written as 
\begin{align}
\omega_P^\mu = \rd x^\mu ~,~~~~~~~~~~~~~~~~~~~~~~~~~\omega_B^\mu &= -\rd x^\nu\partial_\nu\partial^\mu\pi~.
\end{align}
Now something interesting happens. Because we can only build Lagrangians by using these ingredients and higher covariant derivatives, the field $\pi$ must always appear with at least 2 derivatives per field.  Thus the galileons can never be obtained from this construction, since the galileon terms all have fewer than two derivatives per field. While this is puzzling at first, it is possible to show that the $4d$ galileons can nevertheless be realized as {\it Wess-Zumino terms}. 

To see this, we work on the coset space, in which $\pi$ and $\xi^\mu$ are considered as coordinates in addition to those of space-time. We obtain the Lagrangian by integrating a {\it Wess--Zumino} form on the subspace on which $\pi=\pi(x)$ and $\xi^\mu=\xi^\mu(x)$. 
The symmetries on the coset space are generated by the vector fields
\be
C = \partial_\pi~,~~~~~~~~~~~~~~~~~~~
B_\mu = \partial_{\xi^\mu}-x_\mu\partial_\pi~,~~~~~~~~~~~~~~~~~~~
P_\mu = \partial_\mu~.
\label{singlefieldvectors}
\ee
The components of the MC form (\ref{singlefieldMCform}) are the (left) invariant 1-forms on the coset space parametrized by $\{\pi,\xi^\mu,x^\mu\}$, obeying $\pounds_X \omega=0$, where $X$ is any of the vector fields (\ref{singlefieldvectors}) and $\omega$ is any of the forms (\ref{singlefieldMCform}).  

Given this, we construct the Wess--Zumino terms by creating invariant 5-forms by wedging together the 1-forms (\ref{singlefieldMCform}) so that they are invariant under the Lorentz transformations $\mathfrak{so}(3,1)$, and hence are well defined on the coset.  This requires that the Lorentz indices in (\ref{singlefieldMCform}) are contracted using the Lorentz invariant tensors $\eta_{\mu\nu}$ and $\epsilon_{\mu\nu\rho\sigma}$.  

As an example, consider the invariant 5-form
\be 
\omega_1^{\rm wz}=\epsilon_{\mu\nu\rho\sigma}~\omega_ C\wedge\omega_P^\mu\wedge\omega_P^\nu\wedge\omega_P^\rho\wedge\omega_P^\sigma=\epsilon_{\mu\nu\rho\sigma} \rd\pi\wedge \rd x^\mu\wedge \rd x^\nu\wedge\rd x^\rho\wedge \rd x^\sigma \ .
\ee
We may write this as the exterior derivative of a 4-form,
\be 
\omega_1^{\rm wz}=\rd\beta^{\rm wz}_1,\ \ \ \ \ ~~~~~~~~\beta^{\rm wz}_1= \epsilon_{\mu\nu\rho\sigma}\pi\rd x^\mu\wedge \rd x^\nu\wedge\rd x^\rho\wedge \rd x^\sigma \ ,
\ee
which may be pulled back to the space-time manifold $M$ and then integrated, to yield
\be
S^{\rm wz}_1 = \int_M\ \beta^{\rm wz} _1= \int_M\ ~\pi\epsilon_{\mu\nu\rho\sigma}\rd x^\mu\wedge \rd x^\nu\wedge\rd x^\rho\wedge \rd x^\sigma \sim \int \rd^4x~\pi ~.
\ee
This is the tadpole term - the first galileon.  

We continue by considering
\be 
\omega_2^{\rm wz}=\epsilon_{\mu\nu\rho\sigma}~\omega_ C\wedge\omega_B^\mu\wedge\omega_P^\nu\wedge\omega_P^\rho\wedge\omega_P^\sigma=\epsilon_{\mu\nu\rho\sigma} \left(\rd\pi+\xi_\lambda\rd x^\lambda\right)\wedge \rd\xi^\mu \wedge \rd x^\nu\wedge\rd x^\rho\wedge \rd x^\sigma \ ,
\ee
which is also the exterior derivative of a 4-form
\be 
\omega^{\rm wz}_2=\rd\beta^{\rm wz}_2,\ \ \ \ \ \beta^{\rm wz}_2=\epsilon_{\mu\nu\rho\sigma}\left(\pi\rd\xi^\mu-\frac{1}{8}\xi^2\rd x^\mu\right)\wedge \rd x^\nu\wedge\rd x^\rho\wedge \rd x^\sigma \ .
\ee
Pulling back and integrating, we now obtain
\be
S^{\rm wz}_2 = \int_M\ \beta^{\rm wz}_2 = 3! \int \rd^4x\ \left(\pi\partial_\mu\xi^\mu-\frac{1}{2}\xi^2\right) \ .
\ee
Using the Higgs constraint $\xi_\mu=-\partial_\mu\pi$, we obtain the kinetic term - the second galileon,
\be
S^{\rm wz}_2 \sim  \int \rd^4x\ (\partial\pi)^2 \ .
\ee

As a final, and non-trivial, example, let's see how the construction of $\mathcal L_3$ works. Consider
\be 
\omega_3^{\rm wz}=\epsilon_{\mu\nu\rho\sigma}~\omega_ C\wedge\omega_B^\mu\wedge\omega_B^\nu\wedge\omega_P^\rho\wedge\omega_P^\sigma=\epsilon_{\mu\nu\rho\sigma} \left(\rd\pi+\xi_\lambda\rd x^\lambda\right)\wedge \rd\xi^\mu \wedge \rd \xi^\nu\wedge\rd x^\rho\wedge \rd x^\sigma \ ,
\ee
which is the exterior derivative of the 4-form
\be 
\omega^{\rm wz}_3=\rd\beta^{\rm wz}_3,\ \ \ \ \ \beta^{\rm wz}_3= \epsilon_{\mu\nu\rho\sigma}\left (\pi \rd \xi^{\mu}\wedge\rd \xi^{\nu}\wedge \rd x^{\rho}\wedge\rd x^{\sigma}-\frac{1}{3}\xi^{2}\rd\xi^{\mu}\wedge\rd x^{\nu}\wedge\rd x^{\rho}\wedge\rd x^{\sigma}\right )\ .
\ee
Pulling back and integrating gives
\be
S^{\rm wz}_3 = \int_M\ \beta^{\rm wz}_3 =  \int_M\ \rd^{4}x\, \Big[-2\pi\left[(\partial_{\mu}\xi^{\mu})^{2}-\partial_{\mu}\xi^{\nu}\partial_{\nu}\xi^{\mu }\right ]+2\xi_{\alpha}\xi^{\alpha}\partial_{\mu}\xi^{\mu } \Big ] \ ,
\ee
and applying the inverse Higgs constraint $\xi_\mu=-\partial_\mu\pi$, and integrating by parts, we obtain the cubic galileon,
\be
S^{\rm wz}_3 \sim \int_M\ \rd^{4}x\, \square\pi(\partial\pi)^2 \ .
\ee
The higher $4d$ galileons can be obtained from the forms 
\begin{align}
\omega_4^{\rm wz} &=\epsilon_{\mu\nu\rho\sigma}\omega_C\wedge\omega_B^\mu\wedge\omega_B^\nu\wedge\omega_B^\rho\wedge\omega_P^\sigma~,\nonumber\\
\omega_5^{\rm wz} &=  \epsilon_{\mu\nu\rho\sigma}\omega_C\wedge\omega_B^\mu\wedge\omega_B^\nu\wedge\omega_B^\rho\wedge\omega_B^\sigma~.
\end{align}
It follows from this construction that the galileon terms are members of the relative Lie algebra cohomology group 
\be 
H^{5}\left(\mathfrak{Gal}(3+1, 1), \mathfrak{so}(3,1)\right) \ ,
\ee
in a similar way to that in which the usual Wess-Zumino terms can be characterized via de Rham cohomology.

The coset methods discussed in this section provide a fresh perspective on the construction of galileons, and some insight into the mathematical structure underlying them. They are a useful counterpart to the probe brane methods of the previous section.

\section{Summary}
Attempts to modify gravity with an eye to providing explanations for cosmic acceleration have yielded fascinating new theoretical ideas such as the DGP model and massive gravity. Fascinatingly, these theories share similar behaviors in certain limits, described by a scalar field theory -- the galileon theory -- that is interesting in its own right. In this talk I described the original Galileon idea, and then discussed work I have been involved with in recent years to generalize ways to construct galileon-like theories with more complicated structures but similar attractive properties. 

One way is to use the probe brane construction, the embedding of a particular brane in a suitable
ambient space to yield an interesting $4$-dimensional theory. In this case, the resulting symmetries are inherited from combinations of five-dimensional Poincar\'e invariance and brane reparametrization invariance.  In the case of Galileons on a curved background there is a rich story, with maximally symmetric choices for the bulk and brane metrics yielding complicated four dimensional symmetry groups.

Another way to construct galileon-like theories is to use the coset construction to take advantage of the fact that the relevant symmetries are nonlinearly realized. For the simple galileons, the relevant terms arise as Wess-Zumino terms, which can then be categorized through relative Lie algebra cohomology groups. In even more recent work, we have shown~\cite{Goon:2014paa} that massive gravity~\cite{deRham:2010ik,deRham:2010kj} itself arises through the coset construction, although interestingly there the relevant terms are not Wess-Zumino terms, only becoming so in a particular limit.

There remain many open questions for these field theories. In this talk I was only able to cover the basic idea of construction techniques, but it is possible that fascinating new applications are right around the corner. But it is also possible that further investigation of the theoretical constraints on these models may render them inconsistent. Many of us are working hard to try to figure out which of these might be true.

\vspace{5mm}

\noindent
{\bf Acknowledgments:} I would like to thank the organizers of the DISCRETE 2014 conference, at Kings College London, for a very enjoyable meeting and their wonderful hospitality. I am also indebted to my collaborators in the work described here --- Garrett Goon, Austin Joyce, Kurt Hinterbichler and Daniel Wesley. I would also like to thank Melinda Andrews, Justin Khoury and James Stokes for helpful discussions and collaboration. This work was supported by the US Department of Energy grant DE-FG02-95ER40893, and by NASA ATP grant NNX11AI95G.

\bibliographystyle{utphys}

\begin{thebibliography}{99}
\bibitem{Dvali:2000hr}
  G.~R.~Dvali, G.~Gabadadze and M.~Porrati,
  Phys.\ Lett.\ B {\bf 485}, 208 (2000)
  [arXiv:hep-th/0005016].
  
\bibitem{deRham:2010ik} 
  C.~de Rham and G.~Gabadadze,
  Phys.\ Rev.\ D {\bf 82}, 044020 (2010)
  [arXiv:1007.0443 [hep-th]].
  
\bibitem{deRham:2010kj} 
  C.~de Rham, G.~Gabadadze and A.~J.~Tolley,
  Phys.\ Rev.\ Lett.\  {\bf 106}, 231101 (2011)
  [arXiv:1011.1232 [hep-th]].

\bibitem{Nicolis:2008in}
  A.~Nicolis, R.~Rattazzi and E.~Trincherini,
  Phys.\ Rev.\  D {\bf 79}, 064036 (2009)
  [arXiv:0811.2197 [hep-th]].

\bibitem{Luty:2003vm}
  M.~A.~Luty, M.~Porrati and R.~Rattazzi,
  JHEP {\bf 0309}, 029 (2003)
  [arXiv:hep-th/0303116].

\bibitem{Hinterbichler:2010xn} 
  K.~Hinterbichler, M.~Trodden and D.~Wesley,
  Phys.\ Rev.\ D {\bf 82}, 124018 (2010)
  [arXiv:1008.1305 [hep-th]].
 
\bibitem{Burrage:2010cu}
  C.~Burrage, C.~de Rham, D.~Seery and A.~J.~Tolley,
  JCAP {\bf 1101}, 014 (2011)
  [arXiv:1009.2497 [hep-th]].
  
\bibitem{Deffayet:2010zh}
  C.~Deffayet, S.~Deser and G.~Esposito-Farese,
  Phys.\ Rev.\  D {\bf 82}, 061501 (2010)
  [arXiv:1007.5278 [gr-qc]].

\bibitem{Padilla:2010de}
  A.~Padilla, P.~M.~Saffin and S.~Y.~Zhou,
  JHEP {\bf 1012}, 031 (2010)
  [arXiv:1007.5424 [hep-th]].

\bibitem{Padilla:2010ir}
  A.~Padilla, P.~M.~Saffin and S.~Y.~Zhou,
  Phys.\ Rev.\  D {\bf 83}, 045009 (2011)
  [arXiv:1008.0745 [hep-th]].

\bibitem{Padilla:2010tj}
  A.~Padilla, P.~M.~Saffin and S.~Y.~Zhou,
  JHEP {\bf 1101}, 099 (2011)
  [arXiv:1008.3312 [hep-th]].

\bibitem{Zhou:2010di}
  S.~Y.~Zhou,
  Phys.\ Rev.\  D {\bf 83}, 064005 (2011)
  [arXiv:1011.0863 [hep-th]].
  
\bibitem{Vainshtein:1972sx} 
  A.~I.~Vainshtein,
  Phys.\ Lett.\ B {\bf 39}, 393 (1972).
  
\bibitem{Joyce:2014kja} 
  A.~Joyce, B.~Jain, J.~Khoury and M.~Trodden,
  Phys.\ Rept.\  {\bf 568}, 1 (2015)
  [arXiv:1407.0059 [astro-ph.CO]].
  
\bibitem{deRham:2010eu}
  C.~de Rham and A.~J.~Tolley,
  JCAP {\bf 1005}, 015 (2010)
  [arXiv:1003.5917 [hep-th]].
  
\bibitem{Goon:2011qf} 
  G.~Goon, K.~Hinterbichler and M.~Trodden,
  JCAP {\bf 1107}, 017 (2011)
  [arXiv:1103.5745 [hep-th]].
  
\bibitem{Goon:2011uw} 
  G.~Goon, K.~Hinterbichler and M.~Trodden,
  Phys.\ Rev.\ Lett.\  {\bf 106}, 231102 (2011)
  [arXiv:1103.6029 [hep-th]].
  
\bibitem{Goon:2011xf} 
  G.~Goon, K.~Hinterbichler and M.~Trodden,
  JCAP {\bf 1112}, 004 (2011)
  [arXiv:1109.3450 [hep-th]].
  
\bibitem{Goon:2012mu} 
  G.~Goon, K.~Hinterbichler, A.~Joyce and M.~Trodden,
  Phys.\ Lett.\ B {\bf 714}, 115 (2012)
  [arXiv:1201.0015 [hep-th]].
  
\bibitem{Burrage:2011bt} 
  C.~Burrage, C.~de Rham and L.~Heisenberg,
  JCAP {\bf 1105}, 025 (2011)
  [arXiv:1104.0155 [hep-th]].
  
\bibitem{Trodden:2011xh} 
  M.~Trodden and K.~Hinterbichler,
  Class.\ Quant.\ Grav.\  {\bf 28}, 204003 (2011)
  [arXiv:1104.2088 [hep-th]].

\bibitem{Lovelock:1971yv}
  D.~Lovelock,
  J.\ Math.\ Phys.\  {\bf 12}, 498 (1971).

\bibitem{Gibbons:1976ue}
  G.~W.~Gibbons and S.~W.~Hawking,
  Phys.\ Rev.\  D {\bf 15}, 2752 (1977).

\bibitem{York:1972sj}
  J.~W.~York,
  Phys.\ Rev.\ Lett.\  {\bf 28}, 1082 (1972).

\bibitem{Myers:1987yn}
  R.~C.~Myers,
  Phys.\ Rev.\  D {\bf 36}, 392 (1987).

  \bibitem{Coleman:1969sm}
S.~Coleman, J.~Wess and B.~Zumino, 
Phys. Rev. {\bf 177}, 2239 (1969)

\bibitem{Callan:1969sn}
C.~Callan, S.~Coleman, J.~Wess and B.~Zumino, 
Phys. Rev. {\bf 177}, 2247 (1969)

\bibitem{volkov}
 D.~V.~Volkov,
 Sov.\ J.\ Particles and Nuclei\ {\bf 4}, 3 (1973)

\bibitem{xthschool}
 V.~I.~Ogievetsky,
 Proc. of X--th Winter School of Theoretical Physics in Karpacz, Vol. 1, Wroclaw 227 (1974)

\bibitem{inversehiggs}
  E.~A.~Ivanov, V.~I.~Ogievetsky,
  Teor.\ Mat.\ Fiz.\  {\bf 25}, 164-177 (1975).

\bibitem{Weinberg:1968de} 
  S.~Weinberg,
  Phys.\ Rev.\  {\bf 166}, 1568 (1968).

\bibitem{Goon:2014ika} 
  G.~Goon, A.~Joyce and M.~Trodden,
  Phys.\ Rev.\ D {\bf 90}, no. 2, 025022 (2014)
  [arXiv:1405.5532 [hep-th]].
  
\bibitem{Ivanov:1975zq} 
  E.~A.~Ivanov and V.~I.~Ogievetsky,
  Teor.\ Mat.\ Fiz.\  {\bf 25}, 164 (1975).
  
\bibitem{Low:2001bw} 
  I.~Low and A.~V.~Manohar,
  Phys.\ Rev.\ Lett.\  {\bf 88}, 101602 (2002)
  [hep-th/0110285].
  
\bibitem{Goon:2014paa} 
  G.~Goon, K.~Hinterbichler, A.~Joyce and M.~Trodden,
  arXiv:1412.6098 [hep-th].

\end{thebibliography}

\end{document}